**Please note, this is the accepted version of the manuscript.**

**The published journal article is available open access at:**

**You can cite the published journal article as:**

**Liaison, safeguard, and well-being: analyzing the role of social robots during the COVID-19 pandemic**

**Abstract**

We examine the implementation of social robots in real-world settings during the COVID-19 pandemic. In particular, we analyze the areas in which social robots are being adopted, the roles and tasks being fulfilled, and the robot models being implemented. For this, we traced back and analyzed 240 deployment cases with 86 different social robots worldwide that have been adopted since the coronavirus outbreak. We found that social robot adoption during this period was strongly related to the use of this technology for crisis management. The social robots' capacity to perform the roles of liaison to minimize direct contact among humans, safeguard to ensure contagion risk-free environments, and well-being coach to protect mental and physical health, is key to explaining adoption within this context. The results of the study offer a complete overview of social robots' utilization in real life settings during the pandemic.

**1. Introduction**

Physical distancing and isolation measures have been adopted worldwide to contain the spread of the COVID-19 pandemic [1]. However, applying physical distancing is not always a valid option and, when it cannot be guaranteed, may lead to devastating consequences. This is particularly noticeable in the healthcare sector. It is estimated that healthcare workers represented greater than 25% of all diagnosed cases of COVID-19 in some countries during the first wave [2].

Social robots are those specifically designed to interact with humans in human physical environments [3]. In order to create successful human-robot interactions, they follow the behavioral norms expected by humans [4,5] and generally present communicative capabilities through natural language and social cues [6]. Social robots are currently utilized in several areas, including healthcare [7–9], education [10], and the service sector [11,12]. It is generally agreed that these autonomous systems will be increasingly integrated into society over the coming years [13].

The critical role that robots could play in public health during disaster situations has long been highlighted [14–16]. The current health crisis has provided an unprecedented opportunity for social robots to demonstrate their value to assist people in real scenarios [17]. Robots have traditionally been regarded as a useful technology to perform tasks that are unsafe for human beings, such as exploring dangerous environments [18,19]. During the COVID-19 pandemic, robots can act safely in environments that have become momentarily hazardous to human beings. Social robots in particular could be decisive to minimize intrahospital transmissions by performing functions of monitoring or facilitating teleoperation to connect doctors with patients [20]. Additionally, there is empirical evidence that isolation measures, quarantines, and lockdowns have a serious impact on mental health and psychological well-being [21,22]. Social robots could also contribute to palliate mental health issues, reduce isolation, and promote social connectedness during this period [20,23,24]. Overall, social robots are being regarded as a critical technology to create uplifting changes for consumer well-being during the coronavirus outbreak [23].

The present study examines the implementation of social robots in real-world scenarios during the COVID-19 pandemic. Previous works developed within the context of the current health crisis have shown how robot manufacturers and researchers are exploring the potential of robotic systems to overcome the challenges brought by the coronavirus outbreak [25,26]. Interestingly, a study found that the pandemic has positively affected people's perception of social robots as companions and that the feeling of loneliness brought by the measures adopted during this period could drive the purchase of a social robot [27]. Likewise, it has been suggested that the use of robotics during the pandemic might contribute not only to the decline of transmission of the virus by helping maintain social distancing, but also to carrying out tasks more safely and effectively [28].

The study aims to provide an in-depth picture of social robots' deployment in real-world settings during the pandemic. Specifically, the study examines the areas and countries where social robots are being implemented, the roles being fulfilled, the specific tasks being performed, and the robot models being adopted. For this, we traced back and analyzed 240 cases of social robots that have been deployed during the COVID-19 pandemic in real-world scenarios worldwide.

Following [17], we propose three strategic roles that social robots can fulfill to facilitate physical distancing as well as to reduce the drawbacks of isolation during the pandemic:

- Liaison: acting as a link in tasks that would normally entail human-human interaction
- Safeguard: ensuring contagion risk-free environments
- Well-being coaches: providing therapeutic and entertaining functions

All tasks and functions that we identified are classified under these three roles in the results.

## 2. Method

Data collection was performed using a documentary research method, which refers to the study of sources that contain relevant information about the phenomenon that is being examined [29,30]. Following the method adopted in [11], we used a strategic non-probabilistic sample to collect the data, due to the large number of cases and the impossibility to quantify all of them on a global scale.

The keywords "COVID-19", "robots", and "social robots" were used in the search engines to identify the cases. We consulted and examined more than 1,000 documentary sources to extract the largest number possible of cases and retained only the sources that described experiences that met the inclusion criteria. The main sources included mainstream and specialized media outlets, technology blogs, robot developers' and robot suppliers' official websites and social media accounts, research papers and reports, study cases published on corporate websites, national and international robotics organizations' websites and reports, and personal communications through email and social networks. For each case that was identified, as many sources as possible were further localized to contrast information. Data collection was performed between March 2020 and November 2020, which corresponds to the outbreak of the COVID-19 pandemic.

The inclusion of cases was restricted to robots deployed in real-world scenarios and for actual use by end-users. In order to include a case in the study, the following criteria were considered:

- Temporality: Only cases that were deployed or specifically adapted in response to the COVID-19 pandemic were included.
- Sociability: Only robots specifically designed to interact with humans and in human physical environments were considered. Non-social robots implemented for COVID-19 related activities were excluded.
- Embodiment: Only robots with a physical body were included.
- Communication: Only robots that exhibited at least a minimum capability to interact with humans using verbal and/or non-verbal communication, and/or social cues were included.
- Autonomy: Only autonomous and semi-autonomous robots were included.
- Movement and navigation: Only robots with movement capacity through gesturing or/and autonomous navigation were included.
- Geography: Worldwide.

A database was created with all the cases that met the inclusion criteria. Each case was classified according to the following variables: social robot name, manufacturer, sector/area of implementation, country where the deployment took or is taking place, specific tasks performed by the social robot, and organization that implemented the robot.

## 3. Results

The final database (see Supplementary Information – database) contains 240 cases of social robot deployments in real-world scenarios that met the inclusion criteria. Some exceptions to the communication, autonomy, and movement and navigation criteria were made to include some teleoperated machines with robotic appearances identified as robots in the documentary sources. These exceptions were made when the robot represented the only case of deployment that we were able to localize in a country or when the robot performed a function that was not previously included in the database through other robots.

In total, 86 different models of social robots were identified among the 240 cases analyzed. The robots that were most recurrent in these instances are reported in Table 1.

**Table 1**

*Robots with the largest number of deployments*

| Robot Name | Manufacturer | N. of deployments |
|---|---|---|
| Temi | Robotemi | 30 |
| Pepper | Softbank | 14 |
| James | Zorabots | 13 |
| Cruzr | Ubtech | 8 |
| Sanbot Elf | Qihan Technology | 8 |
| Greetbot | OrionStar | 8 |
| Starship | Starship | 7 |

The cases retained in the database were from 41 different countries. The countries in which we identified the largest number of cases are reported in Table 2.

**Table 2**

Countries with the largest number of cases identified in the study

| Country | Number of cases |
|---|---|
| China | 56 |
| The USA | 27 |
| Thailand | 24 |
| Belgium | 14 |
| Japan | 13 |
| Hong Kong | 13 |
| India | 13 |
| South Korea | 11 |

| Spain | 6 |

By location, the largest number of cases was identified in hospitals, followed by nursing homes. Table 3 reports the location of all the deployments included in the study.

**Table 3**

Location by area of the deployments included in the study

| Area | Number of cases |
| --- | --- |
| Hospitals | 122 |
| Nursing Homes | 22 |
| Transportation | 16 |
| Restaurants | 14 |
| Educational Centers | 13 |
| Airports | 10 |
| Hotels | 10 |
| Office Buildings | 10 |
| Outdoor Public Spaces (e.g., parks or streets) | 8 |
| Shopping Malls | 6 |
| Stores | 4 |
| Older adults' homes | 3 |
| Real Estate | 1 |
| Highway Checkpoints | 1 |

The specific functions performed by the robots were classified within the three main roles of liaison, safeguard, and well-being coaches proposed in [17] and are reported in Table 4. In most of the cases that were analyzed (190), the robots performed multiple functions. The capacity to verbally greet people, which generally complemented other functions, was identified in 109 cases.

**Table 4**

Summary of functions performed by social robots classified by role

| **Liaison** |
| --- |
| Delivery |
| Telepresence |
| Monitoring |
| Receptionist |
| Providing Information (general and personalized) |
| Pre-diagnosis |
| **Safeguard** |
| Safety Advice (general and personalized) |
| Protective Measure Enforcement |
| Surveillance and Patrolling |
| Disinfection |
| **Well-being coach** |
| Entertainment |
| Edutainment |
| Companion |
| Medical and Wellbeing Adherence |
| Promotion of Physical Exercise |

The robots that were able to perform the largest number of functions were Temi (Robotemi), with a total of 15 functions across cases and up to 9 in a single case; Pepper (Softbank), with a total of 13 functions across cases and up to 5 in a single case; Cruzr (Ubtech), with a total of 12 functions across cases and up to 7 in a single case; Greetbot (OrionStar), with a total of 12 functions across cases and up to 8 in a single case; Sanbot Elf (Qihan Technology), with a total of 12 functions across cases and up to 6 in a single case; CLOi GuideBot (LG Electronics), with a total of 9 functions across cases and up to 9 in a single case; and Promobot (Promobot), with a total of 9 functions across cases and up to 5 in a single case. Only 16 robots were used for a single function, which was either delivery or disinfection.

A description of the different functions classified by role is presented below.

## 3.1 Liaison

The liaison role included all functions related to using social robots to act as a link between humans to minimize human-human direct contact. Functions associated with the liaison role have become particularly widespread as a result of the pandemic. In particular, 200 of the 240 cases we analyzed and 71 different robot models implemented at least one function that can be related to the liaison role. Interestingly, functions that were not particularly valued before the pandemic such as the possibility to bring food to patients have now become of major importance as they facilitate physical distancing between patients and staff. The functions of delivery, telepresence, monitoring, receptionist, providing information, and pre-diagnosis were identified in relation to this role.

### 3.1.1 Delivery

Delivery is among the functions that present a wider range of implementation as a result of the pandemic. In total, 106 cases and 43 different robot models were identified to carry out transportation or delivery functions either exclusively or in combination with other functions. The delivery function is also prevalent among robots that do not present social features, which were not included in the sample. There are two different groups of delivery robots: indoor and outdoor.

We identify two main types of robots among those that perform indoor object delivery. First, robots with an essentially functional design, such as Pudu (Pudu Robotics), Peanut Delivery Robot (Keenon Robotics), TUG (Aethon), Hospi (Panasonic), or Run (Yunji Technology). These robots are implemented in hospitals, restaurants, and hotels with quarantined guests to transport medication, linens, meals, medical supplies, and documents. Some can also open and close doors, take elevators, transport heavy objects, store medical waste, or disinfect themselves. Second, robots that present more social features and humanoid or semi-humanoid shapes (Fig. 1A). These robots are utilized in reception areas of hospitals, nursing homes, office buildings, shopping malls, restaurants, educational centers, and airports to distribute masks and sanitizer to visitors. Some of these robots were initially designed for other functions, such as waiter robots, and some have just been added a tray and are now utilized for object and meal delivery to patients as an additional function. Examples include Ani (Asimov Robotics), Amy (Pangolin Robotics), Zafi (Propeller Technologies), Sona 2.5 (Club First), Temi (Robotemi), or Sunbot (Siasun Robot and Automation).

Delivery robots in outdoor scenarios have also experienced a notable expansion during the lockdown periods (Fig. 1B). Some specific areas in California [31] and China [32] eased circulation regulations for these robots during the pandemic, which facilitated their adoption. These robots present mostly a functional design. Some social features are included to facilitate communication with pedestrians and package receivers. For instance, lights that suggest eyes are used as a form to indicate that the robot will change direction (Serve - Postmates) or a thumbs-up can be used to order the robot to open or close its doors (Nuro - Nuro). Some delivery robots have additionally been equipped with disinfection functions during the pandemic (e.g., Kiwibot – Kiwibot, Neolix ADV – Neolix). Outdoor delivery robots are used by e-commerce platforms, delivery companies, restaurants, and grocery stores in particular neighborhoods and university campuses as well as to deliver sanitary supplies to healthcare facilities. Kiwibot (Kiwibot), Starship (Starship), RoboPony (Zhen Robotics), Nuro (Nuro), and Neolix ADV (Neolix) are examples of this group of robots.

3.1.2 Telepresence

One of the most widespread functions during the COVID-19 crisis is telepresence (Fig. 1C), with 95 cases and 27 different robot models identified. Robots are equipped with a camera and a screen that communicates between patients in hospitals or residents in nursing homes and medical staff and relatives. Through this function, patients can get a remote diagnosis after interacting with a doctor or request medical assistance. Telepresence robots are found mostly in hospitals and elderly care centers. They have also occasionally been used by schools and universities during the pandemic, for functions such as graduation ceremonies, as well as in hotels and at a highway checkpoint. Some examples of robots with telepresence functions are James (Zorabots), Temi (Robotemi), Vita (InTouch Health), Cutii (CareClever), and Ninja (Chulalongkorn University).

### 3.1.3 Monitoring

In 59 of the cases analyzed, 16 different robot models performed patient monitoring functions. This function is mostly found in hospitals and, occasionally, in nursing homes, and older adults' homes. Monitoring functions include measuring body temperature, blood pressure, oxygen saturation, observing patient routines, and reporting exceptional changes to caregivers. Some of these robots are also equipped with cameras and sensors that allow families or caregivers to remotely monitor the patient, transmit and store information in the patient medical record, broadcast video images of the patients to nurse stations, and keep track of medical adherence. Examples include Sanbot Elf (Qihan Technology), Cruz (Ubtech), James (Zorabots), Cutii (CareClever), Vita (InTouch Health), and Ninja (Chulalongkorn University).

### 3.1.4 Receptionist

A total of 63 cases and 27 different robot models utilized social robots for receptionist-related tasks. In particular, the robots perform patient or customer registration upon arrival, contactless check-ins/outs, book appointments, provide directions, get access to medical records, prepare and print prescriptions, or translate information in different languages. Robots deployed for receptionist tasks tend to present anthropomorphic features, humanoid or semi-humanoid shapes, and verbal interaction capabilities. Most of these robots are found at the reception of hospitals. Some cases were also identified in hotels to check-in/out patients in isolation, in elderly care centers, airports, restaurants, stores, office buildings, and shopping malls, to provide directions and guide visitors and customers to specific areas. Examples of robots that are used for these functions are Cruz (Ubtech), Pepper (Softbank), Mitra and Mitri (Invento Robotics), Mini Ada (Akin Robotics), and Greetbot (OrionStar).

### 3.1.5 Providing Information

Providing information was a function available in 50 of the cases analyzed, with 20 different robot models. The information the robots offer can either be general (28 cases) or personalized (22 cases). General information includes tasks such as informing about schedules or offering promotions in shops. When the patient or customer interacts with the robot and asks for a specific question, the robot can offer personalized information such as the specific room to go for taking a test in a hospital, finding the boarding gate for travelers in airports and accompanying them. Robots develop this function in hospitals, but also stores, hotels, office buildings, restaurants, universities, airports, nursing homes, and older adults' homes. Examples include Medical Service Robot (TMI Robotics), Smart Service Robot (Xiaoben Intelligence), and Pari 2.0 (Paaila Technology).

### 3.1.6 Pre-Diagnosis

COVID-19 pre-diagnosis functions were identified in 31 of the cases, with 16 different robot models. In hospitals, where most cases are found, the robots are able to make a pre-diagnosis, assessing patient symptoms through questionnaires and thermal screenings with the aim of reducing waiting lists. If the patients exhibit some coronavirus-related symptoms, the robot redirects them towards a telepresence consultation or medical staff equipped with personal protective equipment. We also identified some cases of robots that perform COVID-19 pre-diagnosis in a public space (Times Square), a shopping mall, a school, and an airport. The social robots in these spaces also conduct pre-diagnosis through temperature checks and questionnaires. This function works in combination with the function of personalized advice (see *Safety Advice*). Examples of robots used for this task are Mitri and Mitra (Invento Robotics), Promobot (Promobot), and Greetbot (OrionStar).

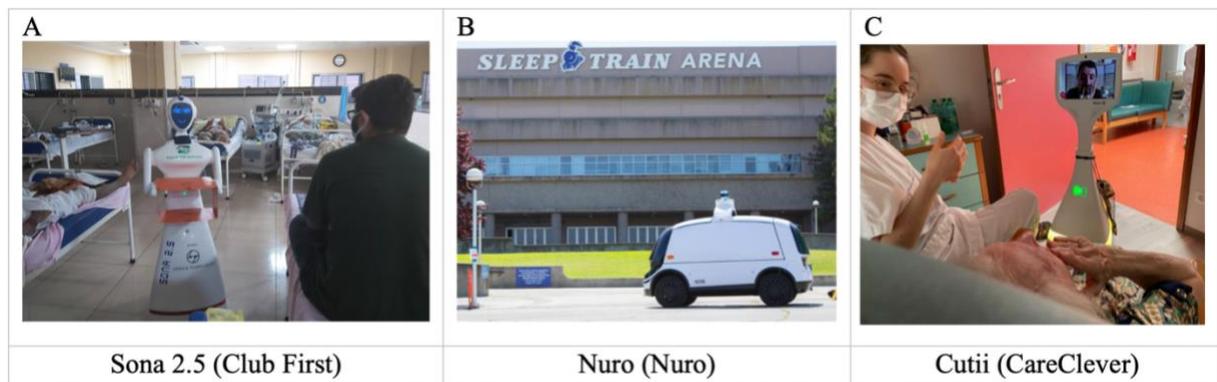

| A | B | C |
| --- | --- | --- |
| Sona 2.5 (Club First) | Nuro (Nuro) | Cutii (CareClever) |

*Figure* 1. Social robots performing functions related to the liaison role[1]

---

[1] Credits: A) Credited to Club First; B) Credited to Nuro; C) Credited to Cutii

## 3.2 Safeguard

The safeguard role includes all functions related to ensuring risk-free environments. Safeguard functions specifically adapted to the situation of the pandemic have appeared during this period. Of 240 cases analyzed, 137 have implemented social robots to perform at least one function that can be associated with the safeguard role, which involves a total of 50 different robot models. The functions of safety advice, protective measure enforcement, surveillance and patrolling, and disinfection were identified in relation to this role.

### 3.2.1 Safety Advice

Of the sample, 77 deployments and 26 different robot models were used to provide safety advice during the pandemic. We distinguish two sub-types of functions within this group: general (53 cases) and personalized (24 cases) advice. General advice includes promoting safety measures and making recommendations about how to behave in specific places. This function works essentially as a public announcement. Robots offering general advice are found in hospitals, and to a lesser extent, in airports, public places, shopping malls, educational centers, office buildings, grocery stores, restaurants, checkpoint highways, and hotels. Examples include Sanbot Elf (Qihan Technology), Smart Service Robot (Xiaoben Intelligence), and Ani (Asimov Robotics). Personalized advice can be offered by the robot after completing a pre-diagnosis function (see *Pre-diagnosis*). If symptoms are identified, the robot advises seeking medical assistance. These robots are principally located in hospitals, but some deployments were also found in airports, office buildings, schools, stores, and shopping malls. Cruzr (Ubtech), Promobot (Promobot), Mitra and Mitri (Invento Robotics), Medical Service Robot (TMI Robotics), and Pepper (Softbank) are some examples of robots that can offer personalized advice.

### 3.2.2 Protective Measure Enforcement

Tasks related to enforcing COVID-19 protective measures were identified in 67 cases and 25 robot models. Rather than informing, these robots are particularly deployed to enforce safety measures either in one-to-one security scans or among crowds. These robots are able to detect if people wear masks, maintain physical distance, or have a fever by measuring temperature (Fig. 2A). When there is a safety concern regarding these aspects, the robot can take actions such as asking the person to wear a mask, keep physical distance, or ban entrance to an indoor space. These robots are principally found at the entrance door of hospitals, nursing homes, shopping malls, grocery stores, hotels, restaurants, educational centers, and office buildings (Fig. 2B). They can also be found navigating airports, outdoor public spaces, or highway checkpoints. Examples include Cruzr (Ubtech), Spot (Boston Dynamics), Dinsow Series (CT Asia Robotics), Aimbot (Ubtech), Temi (Robotemi), Promobot (Promobot), Mitri (Invento Robotics), or AMY Service Robot (AMY Robotics).

### 3.2.3 Surveillance and Patrolling

Surveillance functions aimed at preventing the spread of the virus were identified in 45 cases and 20 different robot models. Robots can be equipped with regular and thermal cameras, lights, speakers, microphones, real-time transmission, and camera recording. These robots are used for patrolling outdoors or indoors. They are either connected to security staff who can communicate through it or programmed to take action when a risk is detected (see *Protective Measure Enforcement*). In public places, they can be found in parks, parking lots, public transportation stations, and congested areas in cities. Robots typically deployed in outdoor spaces are ASR (KnightScope), O-R3 (Otsaw Digital), Atris (Ubtech), PGuard (Enova Robotics), Outdoor Inspection Robot (Hangzhou Guochen Robot Technology), or Spot (Boston Dynamics). Also, they are deployed in buildings such as hospitals, educational centers, shopping malls, stores, airports, nursing homes, restaurants, and offices, with robots such as Aimbot (Ubtech) or Temi (Robotemi).

### 3.2.4 Disinfection

Disinfection robots (Fig. 2C) have experienced a notable expansion during the pandemic. Most of these robots tend to be industrial (and therefore not included in the database). Forty cases of disinfection robots and 17 different robot models that present at least a minimal social component were identified. We distinguish two sub-types. First, disinfection robots specifically built for this purpose (13 cases). They have a functional design and present only minimal social features, such as verbal warnings. These robots are mostly found in hospitals and some also in airports. Examples are the UVD Robot (Blue Ocean Robotics) and the Intelligent Disinfection Robot (TMI Robotics). Second, robots that have added this function during the pandemic but develop other functions as well, such as surveillance or transportation (27 cases). Principally, they incorporate a spray to potentially disinfect during navigation. These robots are found in hospitals, educational centers, airports, restaurants, public transport stations, and office buildings. Examples are Temi (Robotemi), Atris and Aimbot (Ubtech), KiwiBot (KiwiBot), and Neolix ADV (Neolix).

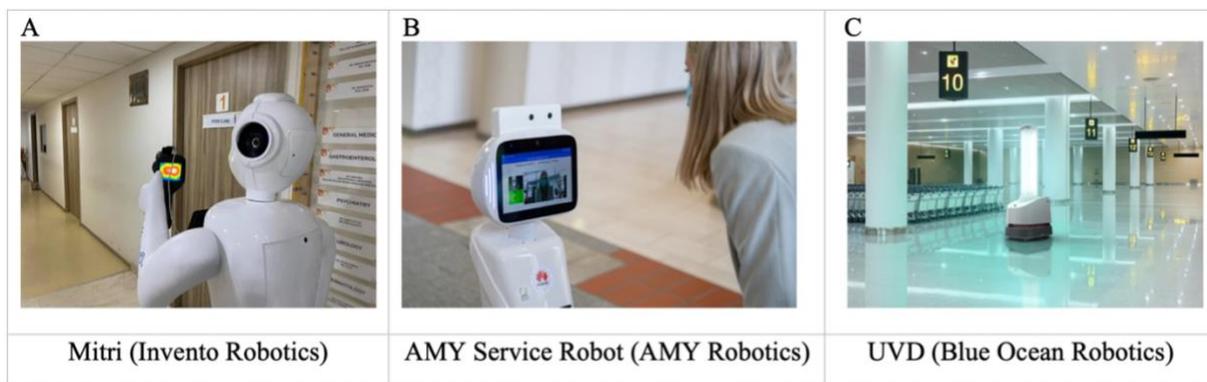

| A | B | C |
|---|---|---|
| Mitri (Invento Robotics) | AMY Service Robot (AMY Robotics) | UVD (Blue Ocean Robotics) |

*Figure* 2. Social robots performing functions related to the safeguard role[2]

### 3.3 Well-being Coach

The well-being coach role includes all functions related to preserving physical and psychological well-being. A total of 69 cases, which involved 25 different robot models, included at least one function connected with this role. Overall, the potential of social robots to act as well-being coaches during the sanitary crisis remains underdeveloped as the functions belonging to this role have rarely been specifically adapted to the particular circumstances of the pandemic. The functions of entertainment and edutainment, companion, medical and well-being adherence, and promotion of physical exercise were identified in relation to the well-being coach role.

---

[2] Credits: A) Credited to Invento Robotics; B) Credited to Huawei Austria; C) Credited to Blue Ocean Robotics

### 3.3.1 Entertainment and Edutainment

Robots for entertainment (47 cases and 21 different robot models) and edutainment (eight cases and four different robot models) present a strong social component and are generally humanoid or semi-humanoid. Entertainment functions include dancing, singing, playing games, taking and sharing photos with contacts, navigating on the internet, reading news, displaying photos and videos on screen, or telling jokes and stories, among others (Fig. 3A). Edutainment functions involve brain-training exercises like dictation, reading, or playing memory games. Most cases are deployed in nursing homes, older adults' homes, and hospitals. Some scattered cases of deployment were located in shopping malls, airports, hotels, office buildings, and restaurants. Examples of robots deployed for these purposes are Cruzr (Ubtech), Sanbot Elf (Qihan Technology), Pepper (Softbank), ElliQ (Intuition Robotics), and Temi (Robotemi).

### 3.3.2 Companion

Forty-three cases and 16 different robot models were identified to offer company and comfort. These robots provide motivational conversation through verbal interaction and by expressing and interpreting emotions. They can also have physical contact with the patient or walk with them. These interactions contribute to emotional support and help reduce anxiety during isolation periods. More than half of these deployments were found in hospitals and a significant number also in nursing homes (Fig. 3B). There were a few deployments in older adults' homes, hotels with quarantined guests, at educational centers to emotionally support students during the outbreak, or in office buildings to reduce stress levels of employees. Examples are Sanbot Elf (Qihan Technology), XR-1 (INNFOS), Pepper (Softbank), Temi (Robotemi), ElliQ (Intuition Robotics), Cutii (CareClever), and Dinsow mini (CT Asia Robotics).

### 3.3.3 Medical and Well-being Adherence

Robots that perform medical and well-being adherence functions by reminding to take prescribed medication, appointments, or engaging in healthy habits were identified in eight cases, with five different robot models. These robots generally perform other functions such as monitoring, entertainment, providing general information like news or weather, or telepresence (see these functions above). Deployments are found principally in nursing homes, older adults' homes, and hospitals (Fig. 3C). Examples of robots used for adherence are Pepper (Softbank), Temi (Robotemi), ElliQ (Intuition Robotics), Dasomi (Wonderful Platform), and Dinsow mini (CT Asia Robotics).

### 3.3.4. Promotion of Physical Exercise

Social robots that instruct physical exercise, relaxation techniques, dancing, and yoga to patients, residents, or staff in hospitals and nursing homes during the pandemic were identified in eight cases and five robot models. These robots either perform the movements with their own body or show them on a screen. Examples of robots used for this function are Sanbot Elf (Qihan Technology), XR-1 (INNFOS), Pepper (Softbank), Temi (Robotemi), and Cutii (CareClever).

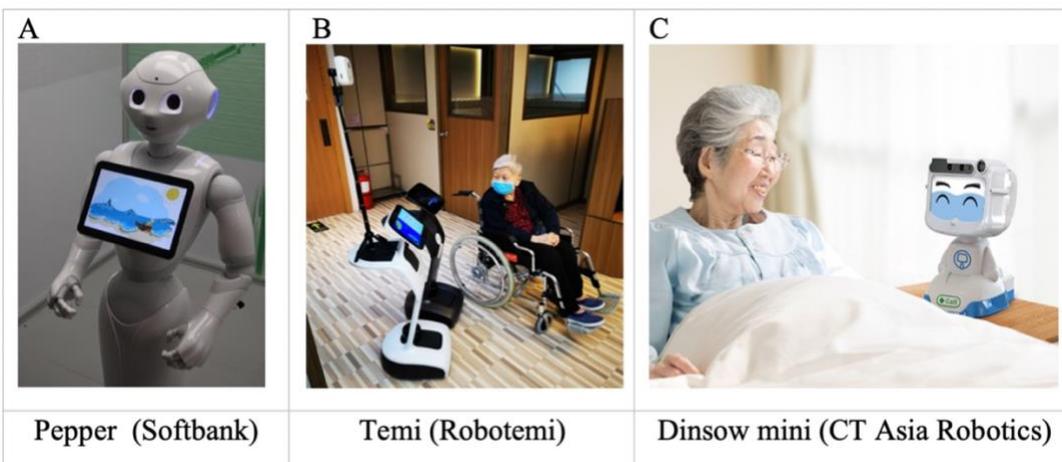

| A | B | C |
| Pepper (Softbank) | Temi (Robotemi) | Dinsow mini (CT Asia Robotics) |

*Figure* 3. Social robots performing functions related to the well-being coach role[3]

---

[3] Credits: A) Luke McKernan, 2017, "Pepper", CC BY-SA 2.0; B) Credited to Temi HK; C) Credited to CT Asia Robotics CO., LTD.

## 4. Discussion

The present study traced back and analyzed 240 deployment cases with 86 different social robots worldwide that were implemented during the COVID-19 pandemic outbreak. We identified experiences in 41 different countries. China was the country in which we found the largest number of experiences (56 cases). Regarding area, hospitals were the place where most deployments took place (122 cases). Concerning robot models, Temi from Robotemi (30 cases), followed by Pepper from Softbank (14 cases), and James from Zorabots (13 cases) were the robots most frequently deployed.

Following [17], we proposed three main roles that social robots fulfill during the pandemic: liaison, safeguard, and well-being coach. The social robots' capacity to act as a liaison, safeguard, and well-being coach is directly associated with the need of facilitating physical distance and palliating the effects of isolation. In particular, the liaison role refers to social robots acting as a link between humans in tasks that would normally require human-human direct contact. This role was involved in 83% of the cases. The safeguard role includes functions related to ensuring risk-free environments. It appeared in 57% of the cases. The well-being coach role is related to preserving physical and psychological well-being. This role was identified in 29% of the cases.

We used this role categorization to classify the different tasks and functions that we identified in our analysis. The functions of delivery, telepresence, monitoring, receptionist, providing information, and pre-diagnosis were found in relation to the liaison role. Safety advice, protective measure enforcement, surveillance and patrolling, and disinfection were identified in relation to the safeguard role. Finally, entertainment and edutainment, companion, medical and well-being adherence, and promotion of physical exercise were found in relation to the well-being coach role. Delivery (106 cases), safety advice (77 cases), and entertainment (47 cases) were the functions identified in more experiences for each role, respectively. Overall, delivery (106 cases), telepresence (95 cases), safety advice (77 cases), and protective measure enforcement (67 cases) were the most recurrent functions.

Social robots are developing a particularly relevant role in hospitals by supporting staff. They contribute to reducing the workload and ensuring the safety of the healthcare personnel performing functions associated with receptionist, pre-diagnosis, food delivery, telepresence, and monitoring, among others. They have also gained acceptance as social companions to palliate the effects of isolation, especially in nursing homes and older adults' homes. In outdoor and public spaces, these robots have also found an opportunity to develop functions aimed at minimizing the spread of the virus, such as disinfection, delivery, surveillance, or protective measure enforcement.

An aspect to highlight is that most functions that we identified were technically available before the pandemic. However, the relevance of these functions is now boosted due to the needs brought by the sanitary crisis and they have been specifically adapted to a situation that demands physical distancing and isolation. This adaptation is particularly visible in the functions associated with the roles of liaison and safeguard. Robots are specifically programmed to detect aspects such as whether people wear masks or keep a physical distance. Robots have also been equipped with technologies that are meaningful during the outbreak such as temperature measurement or surface disinfection. Some functions such as carrying food in hospitals, which were previously non-essential, have now gained more relevance as they help minimize human-human contact. Finally, the functions associated with well-being have not experienced a major adaptation. There is clear room for improvement in the features associated with providing psychological aid to support patients under isolation.

The companies that develop social robots have quickly spotted the opportunity presented to this industry as a result of the health emergency and have implemented marketing strategies to promote their robots accordingly. Several companies have offered their robots at zero cost to hospitals, nursing homes, and elderly people living alone to help fight the pandemic and promote their robots for this purpose. These actions have contributed to generating publicity. Some public research initiatives have also promoted social robot deployments to minimize the drawbacks of the pandemic. Additionally, some countries where robotization is still not largely implemented, such as Vietnam, Senegal, or Malaysia, have developed their own robots to support healthcare workers during the coronavirus outbreak. In general, press releases tend to be positive towards initiatives of social robot deployments for fighting the pandemic and suggest that social robots have gained larger social acceptance during this crisis.

Our study has some limitations. First, the method we utilized presents some flaws. There are several factors, such as language barriers, the ease of access and availability of online information from certain experiences, or the media coverage, that determined what cases we were able to identify and analyze. Also, the results we provide need to be interpreted with caution as our sample is not representative. Additionally, the study we presented is descriptive in nature and therefore we were not able to provide data such as the performance quality of social robots within these settings. Finally, we documented social robot deployment from data collected between March and November of 2020. It would be interesting that future studies performed a longitudinal analysis with additional data points (before and after) to further examine whether the pandemic has boosted social robot adoption.

To conclude, our study examined the areas and countries where social robots are being deployed, the roles being fulfilled, and the tasks being performed. It also provides valuable information regarding the robot models being adopted during the pandemic. The type of roles and functions identified in our analysis suggests that social robot adoption during the coronavirus outbreak is strongly linked to the use of this technology for crisis management, particularly, to facilitate physical distancing and soothe psychological distress. We expect that the results of the study will contribute to providing a complete overview of social robot adoption in real life scenarios during the pandemic and will also encourage future research that can complement our findings.

## 5. Funding

L.A.-F. is supported by the Ramón y Cajal Fellowship Program (ref. RYC-2016-19770), funded by Agencia Estatal de Investigación, Ministerio de Ciencia, Innovación y Universidades, and the European Social Fund.